\documentclass[aps,prb,twocolumn,groupedaddress,showkeys,amsmath,amssymb]{revtex4}

\usepackage{longtable}
\usepackage{graphicx}
\usepackage{dcolumn}
\usepackage{bm}

\begin{document}

\preprint{Submitted to the Journal of Applied Physics}

\title{Point defect energetics in silicon using the LDA+U method}


\author{Panchapakesan Ramanarayanan}
\email[]{panchram@stanford.edu}
\affiliation{Department of Mechanical Engineering, Stanford University,
  California 94305, USA}
\author{Renat F. Sabirianov}
\email[]{rsabirianov@mail.unomaha.edu}
\affiliation{Department of Physics, University of Nebraska, Omaha,
Nebraska 68182, USA}
\author{Kyeongjae Cho}
\email[]{kjcho@stanford.edu}
\affiliation{Department of Mechanical Engineering, Stanford University,
  Stanford, California 94305, USA}


\nopagebreak

\begin{abstract}
We present the first principles results of point defect
energetics in silicon calculated using the LDA+U method: a
Hubbard type on-site interaction added to the local density approximation
(LDA).  The on-site Coulomb and exchange parameters were tuned to match
the experimental band gap in Si.  The relaxed configuration was
obtained using the LDA; LDA+U was used only to calculate the energies.
Our values of point defect energetics are in very good agreement with both
recent experimental results and quantum Monte Carlo (QMC) calculations.
\end{abstract}


\maketitle


\section{\label{sec:intro}Introduction}
Modeling techniques empower the investigator with the capability to
explore into the details of natural phenomena with great dexterity, aptly
complementing their experimental counterparts.  Hierarchical multiscale
modeling (HMM) is one such technique.  We have recently investigated self
diffusion in silicon-germanium alloys in considerable detail by developing
a database of first principles energetics and using the database to
perform kinetic Monte Carlo simulations: a typical HMM scheme
\cite{rama03a}.  The exponential growth in computational 
processing power makes such schemes more viable; the explosive growth in
nanotechnology provides technologically relevant platforms that are
amenable to such schemes. 

Understandably, the accuracy of the results such as those presented in
Ref.~\onlinecite{rama03a} hinges to a great extent on the correctness of 
the energetics database.  In that study, we used the popular local density
approximation (LDA) to develop the energetics database.  However, the
(approximately 1 eV) discrepancy \cite{souz98a} between the theoretical activation energy for Si
self diffusion computed using the LDA (or the generalized gradient
approximation (GGA)) and the experimental values was an important reason
that precluded us from making direct comparisons of our results with
experimental studies of self diffusion in SiGe alloys such as those in
Refs.~\onlinecite{zang01a, stro02a}.

Our present work is directed essentially at finding a means of developing
a reliable point defect energetics database (like in
Ref.~\onlinecite{rama03a}) using an as parameter free a technique as
possible.  (While considerable progress has been made on techniques like
the quantum Monte Carlo (QMC) method, it is beyond the current computational
capacity to use it to develop a detailed energetics database as required
for our HMM scheme.)  We focus on the activation energy for self-diffusion
in pure Si. For self diffusion in Si, we consider the vacancy mechanism,
the interstitialcy mechanism, and the concerted exchange mechanism. These
three have been shown to be the most likely diffusion mechanisms in
silicon \cite{ural98a,ural99a}. For the interstitialcy mechanism, we
consider the hexagonal to split $\langle 110 \rangle$ mechanism
as this has shown to be the most likely mechanism \cite{need99a}. 
 
This article is organized as follows: In Sec. \ref{sec:method}, we provide the details
of the LDA and the LDA+U calculations.  In Sec. \ref{sec:results}, we present the results
of our calculations.  We provide an intuitively appealing explanation for
the underperformance of the LDA.  We then provide a brief overview of the
LDA+U technique and discuss the results obtained using this method for the
different diffusion mechanisms. We summarize our article in Sec. \ref{sec:summary}.

\section{\label{sec:method}Method}
Self-consistent electronic structure calculations were performed using
the plane-wave pseudopotential code VASP \cite{VASP1,VASP2,VASP3,VASP4}
with the projector augmented-wave (PAW) potentials \cite{VASPAW1,VASPAW2} at a kinetic energy cutoff
of 20 Rydberg.  A 64 atom supercell 
was used.  Electronic minimization was carried out to a tolerance of 
$2.7 \times 10^{-5}$ eV
and structures were relaxed until the maximum force on any atom
was less than 0.015 eV/{\r{A}}.  Saddle point configuration for the hexagonal
to split $\langle 110 \rangle$ interstitialcy mechanism was determined using the nudged
elastic band method \cite{NEB}.  Structural relaxation was performed in two stages:
initially with a $2^3$ Monkhorst-Pack \cite{monk76a} {\bf k}-point sampling followed
by a $6^3$ Monkhorst-Pack \cite{monk76a} {\bf k}-point sampling.  Energy calculations were done
using tetrahedron method with a $6 \times 6 \times 6$ division of Brillouin zone.
The lattice constant of systems
containing point defects (vacancy or interstitial) was determined by
fitting the total energy versus the supercell volume to Murnaghan's
equation of state\footnote{The Murnaghan's equation of state \cite{murn44a} is a
relation between the energy $E$ and the volume $V$
of the supercell and has the following form:
$E = E_0 + [(V-V_0)/b] - \{[(V^{1-a} - V_0^{1-a})V_0^a]/[(1-a)b]\}$.
$E_0$, $V_0$, $a$, and $b$ are the fitting parameters.}.

The plane-wave psedopotential code VASP \cite{VASP1,VASP2,VASP3,VASP4} also has
the option to perform calculations by the LDA+U method. (A brief overview of the
method itself will be presented in Sec. \ref{sec:results}. Here we merely give the details of the
parameters used in our calculations.) We have used the rotationally invariant LDA+U scheme 
according to Liechtenstein et al. \cite{liec95a}. The on-site interactions were added
for the {\em p}-orbitals. The effective on site Coulomb interaction parameter ($U$)
was set to 0 eV and the effective on site exchange interaction parameter ($J$) was 
set to 4 eV. (We provide justification for this choice in Sec. \ref{sec:results}.) The LDA
was used to perform structural relaxation; the LDA+U was used only to
perform energetics calculation on the structure obtained using the LDA. We
adopted this approach\footnote{We would like to point out that 
Leung {\em et. al.} \cite{leun99a} take a 
similar approach: they calculate their structure using LDA and use QMC
to compute the energetics.}
because the LDA gives structutal properties
(eg. lattice constant, see Table \ref{tab:resultstab}) that are closer to experimental values
than the LDA+U.

\section{\label{sec:results}Results and Discussion}
Table \ref{tab:resultstab} summarizes the results of both the LDA and LDA+U calculations.
It also contains experimental values and values from quantum Monte Carlo
simulations where available.

There is consensus on the values obtained within the LDA by different
theoretical groups \cite{prob03a,pusk98a,need99a}. It can be seen that the
LDA gives activation energy that is systematically lower than the experimental
values. 

\subsection{\label{sec:ldaresult} Local density approximation}

One of the main reasons for the underperformance of the LDA is that it
tends to overbind.  Because of this, the dangling bonds that are created
due to the defects, tend to bind with each other and with other orbitals.
Because of this overbinding, the energy of the system with the defect is
lower than it should be in reality.  Thus the difference between the
energy of the pure crystal and the one with the defects (which is in fact
the formation or the migration energy, depending on the situation) is
lower than it should be in reality.  The reason for the LDA's overbinding
can be traced to the spherical nature of the exchange hole. This leads to
a larger negative value of the exchange energy and hence lowers the energy
of the system much more than it should be in reality \cite{koch00a}. 

\begin{figure}
\includegraphics[width=8.5cm]{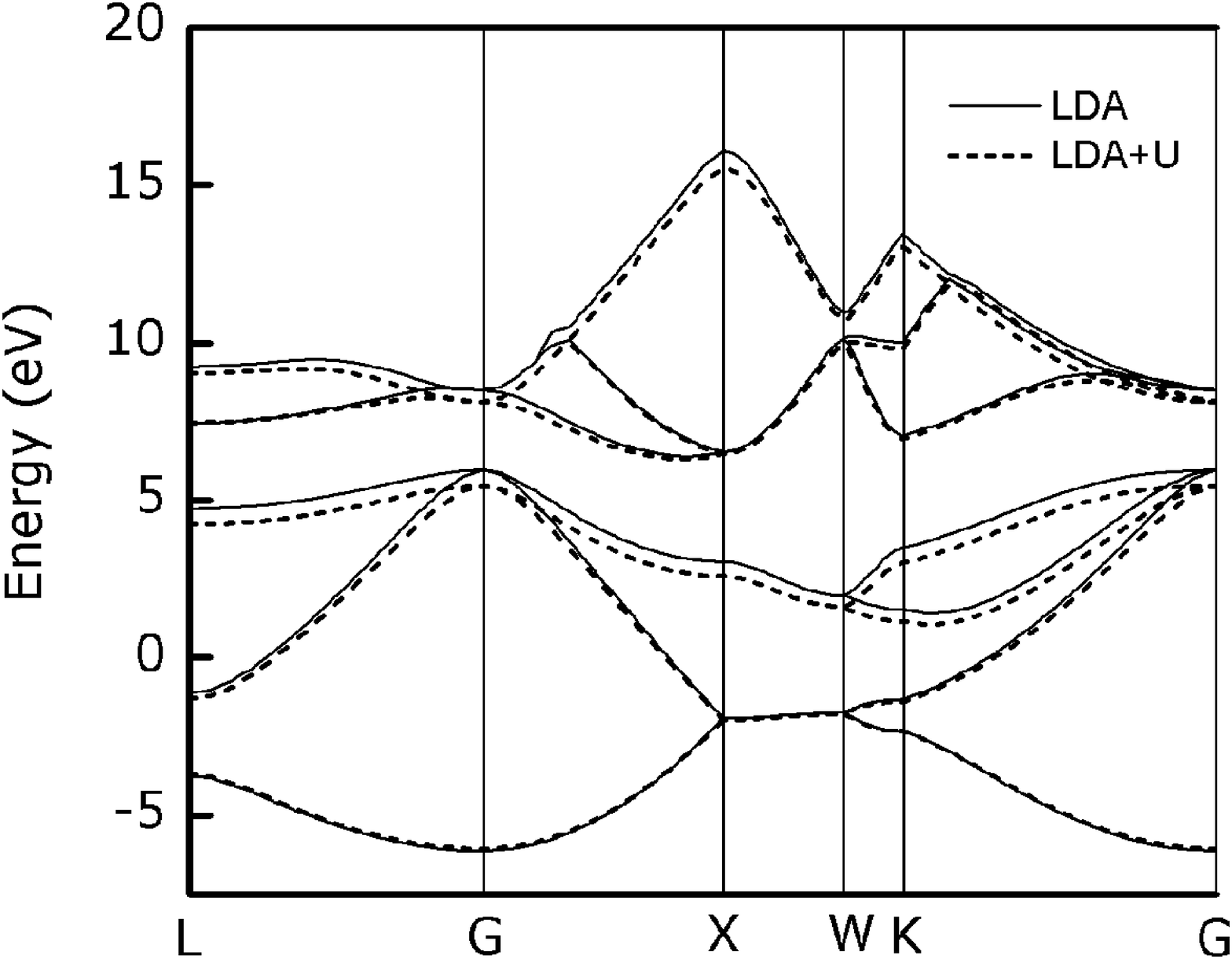}
\caption{\label{fig:si2} {\small Si band structure computed using the LDA
(solid line) and the LDA+U (dashed line) techniques. The (indirect) band
gap using LDA (LDA+U) is 0.56 (0.95) eV. }}
\end{figure}

A related reason for the overbinding nature is the well-known
underestimation of the band gap by LDA.  From our LDA based calculations,
we find that, for example, the Si with a vacancy has a band gap of only
0.88 eV in contrast to experimental 
values of 1.12 eV.  (We note that the band gap in pure Si from LDA
calculations is even less: 0.56 eV (Figs. \ref{fig:si2},
\ref{fig:si64}). The system with 
the impurity has a 
higher band gap because of the interaction of the defect with the other
atoms in the lattice. This can be seen from a simple tight binding
analysis. We consider the band gap of the
system with the defect because that is the environment that the defect
``perceives''.)  Because of the small band gap, the defect states that
are (usually) created in the band gap tend to interact more strongly with the
valence states. This therefore leads to overbinding too. The band
structure and the density of states of the system with a vacancy,
hexagonal interstitial and split $\langle 110 \rangle$ interstitial are
shown respectively in Figs. \ref{fig:si63}, \ref{fig:si65h}, and
\ref{fig:si65s}.
 
\newcommand{\rb}[1]{\raisebox{1.5ex}[0pt]{#1}}
\begin{table}
\caption{\label{tab:resultstab}
  Theoretical results and experimental values for the following
  properties[symbol](units) of silicon: 
  lattice constant[$L_0$]({\r{A}}),
  vacancy formation volume$^{\dagger}$[$V^f_V$]({\r{A}}$^3$),
  vacancy formation energy[$E^f_V$](eV),
  vacancy migration energy[$E^m_V$](eV),
  interstitial formation volume$^{\dagger\dagger}$[$V^f_I$]({\r{A}}$^3$),
  hexagonal interstitial formation energy[$E^f_{HI}$](eV),
  split $\langle 110 \rangle$ interstitial formation energy[$E^f_{SI}$](eV),
  interstitial migration energy[$E^m_I$](eV),
  concerted exchange migration energy[$E^m_{CX}$](eV). Superscripts indicate
  reference numbers. Quantum Monte Carlo results are from Ref. \onlinecite{leun99a}.}
\begin{ruledtabular}
\begin{tabular}{lcccc}
Prop. & LDA & Expt. & QMC & LDA+U\\
\hline
$L_0$      & 5.4   & 5.43 \cite{kitt96a}                   & ---  & 5.27 \\
$V^f_V$    & 4.41  & ---                                   & ---  & ---  \\
$E^f_V$    & 3.49  &                                       & ---  & 4.59 \\
$E^m_V$    & 0.03  & \rb{4.86 \cite{ural99b}$^{\ddagger}$} & ---  & 0.42 \\
$V^f_I$    & $-$2.77 & ---                                   & ---  & ---  \\
$E^f_{HI}$ & 3.37  &                                       & 4.82 & 4.61 \\
$E^f_{SI}$ & 3.34  & 4.68 \cite{ural99b}$^{\ddagger}$      & 4.96 & 4.70 \\
$E^m_I$    & 0.18  &                                       & ---  & 0.44 \\
$E^m_{CX}$ & 4.56  &                                       & 5.78 & 5.82 \\
\end{tabular}
$^{\dagger}$$\;V^f_V$ = Relaxation volume $+$ Atomic volume\\
$^{\dagger\dagger}$$\;V^f_I$ = Relaxation volume $-$ Atomic volume\\
$^{\ddagger}$ These values are the experimental estimates of the activation energy
i.e., the sum of formation and migration energies.
\end{ruledtabular}
\end{table}

There are a couple of interrelated reasons for the LDA's underestimation
of the band gap.  The discontinuity of the one-electron potential for
localized states, which is a characterisitc of the exact density functional, is absent
in the LDA as was shown by Perdew {\em et. al.} \cite{perd82a}. Because
the band gap can be expressed as
\begin{eqnarray}
E_{gap} = E[N+1]+E[N-1]-2E[N]
\end{eqnarray}
where $E[x]$ is the energy of the system with $x$ electrons,
the absence of the discontinuity of one-electron potential in the LDA
causes the LDA to give an incorrect band gap.  The second reason for the
underestimation of band 
gap within LDA is the absence of self interaction correction.  The third
reason has to do with the Kohn-Sham approach \cite{kohn65a} itself.  In
the Kohn-Sham theory, there is no direct relationship between the orbital
energies and the ionization energies (unlike that in Hartree-Fock theory)
except for the highest occupied orbital.
              
\subsection{\label{sec:ldauresult}Local density approximation with Hubbard type correction: LDA+U}

The LDA+U method has been widely applied to metallic systems containing
localized electrons.  We refer the reader to the review article on LDA+U
by Anisimov {\em et. al.} \cite{anis97a}.  Here we provide a brief
explanation of 
the technique along the lines of ``rectifying'' the deficiencies of the
local density approximation that were pointed out in Sec. \ref{sec:ldaresult}
. 
The basic idea of the LDA+U method is to redistribute the
electron-electron interaction.  Because, as we explained in Sec. \ref{sec:ldaresult},
overbinding of the localized electrons associated with the dangling bonds
is a primary reason for the underperformance of the LDA, a direct way of
rectifying this deficiency would be to add an on-site Coulomb repulsion term
while subtracting an average overall electron-electron interaction.
\begin{widetext}
\begin{eqnarray}
H_{LDA+U} = H_{LDA} 
          + \sum_{i=1}^{N}\sum_{j=1;j\neq i}^{N}U_{ij}
        \int |\psi_i(\bm{r}_1)|^2\;d\bm{r}_1
	\int |\psi_j(\bm{r}_2)|^2\;d\bm{r}_2
        - U\frac{N(N-1)}{2}
\end{eqnarray}
\end{widetext}
where $N=\sum_{i=1}^{N}\int|\psi_i(\bm{r})|^2\;d\bm{r}$ corresponds to the
total number of electrons in the system. The parameter $U$ is adjusted by
trial and error so that the resulting band gap equals the experimentally
observed band gap of the system in question.  The above technique, which
only has the Coulomb term, is
generalized to include the exchange term as well.  We refer the reader to
Refs. \onlinecite{liec95a,solo94a} for more details. The $UN(N-1)/2$ introduces the
discontinuity of the of the one-electron potential. We wish to point out
that the LDA+U does not correct for the self interaction.

As mentioned in Sec. \ref{sec:method}, we have used the scheme as implemented in VASP for
our calculations. 
Ideally, the Coulomb ($U$) and exchange ($J$) interaction parameters would be chosen
so that the band gap for the bulk system would 
be equal to the experimentally observed band gap.  However, because our
calculations are based on a supercell geometry, there is an inevitable but
unphysical interaction between the defect and the matrix.  This alters the
band gap 
from that in the bulk. However, because the system ``sees'' only this band
gap, we have chosen $U$ and $J$ so that the band gap in the system with
the defect is close to the experimentally observed band gap.  This, of
course, would mean that the parameter be tuned for each type of defect.
However, for the sake of demonstration, we have worked with a single set of
parameters. 

\begin{figure*}
\includegraphics[width=17cm]{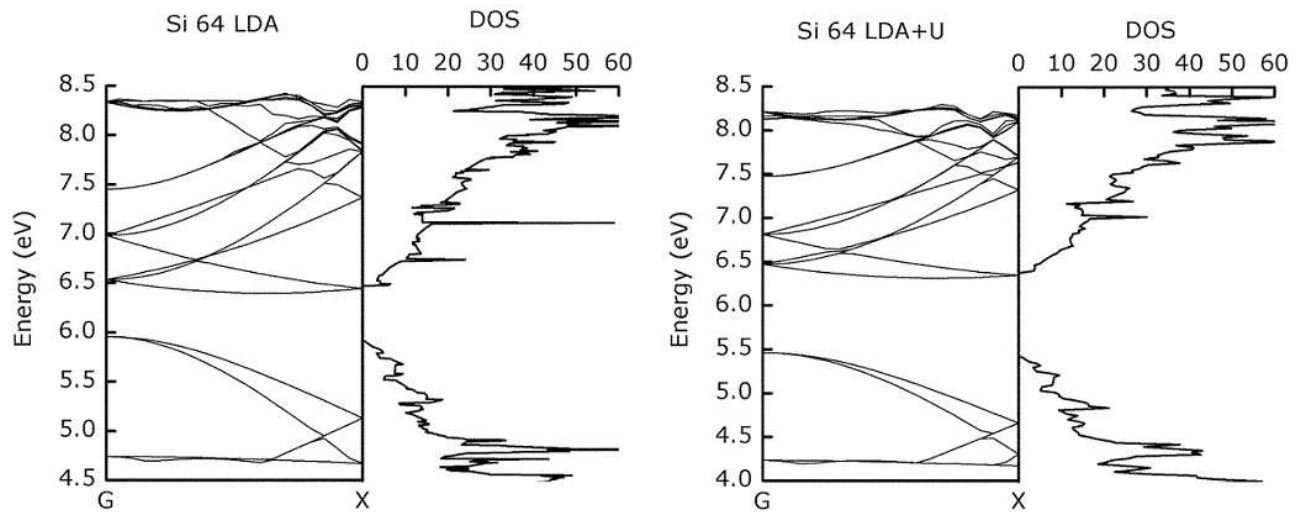}
\caption{\label{fig:si64}{\small Band structure and density of states of
bulk Si computed using the LDA (left) and the LDA+U
(right) methods. This figure is for reference to compare with similar
figures for systems with defects.}}
\end{figure*}

We find that the defect formation and migration energies
increase almost linearly with the difference: $U-J$. The best agreement
between the
above energies and the experimental data occurs when the calculated band
gap corresponds to the experimental one. (This occurs when we choose $U=0$
and $J=4eV$ as mentioned in Sec. \ref{sec:method}.) This supports our argument that the
correction of the density functional for the bandgap (discontinuity of the
potential on the number of electrons) should improve the results. Table \ref{tab:resultstab}
summarizes our energetics 
calculations. Figs. \ref{fig:si2}-\ref{fig:si65s} show band structures and
densities of states of the various systems considered.

\subsection{\label{sec:defstresult}Defect structure, symmetry, states and energy}
\subsubsection{\label{sec:vacresult}Vacancy} 
\begin{figure*}
\includegraphics[width=17cm]{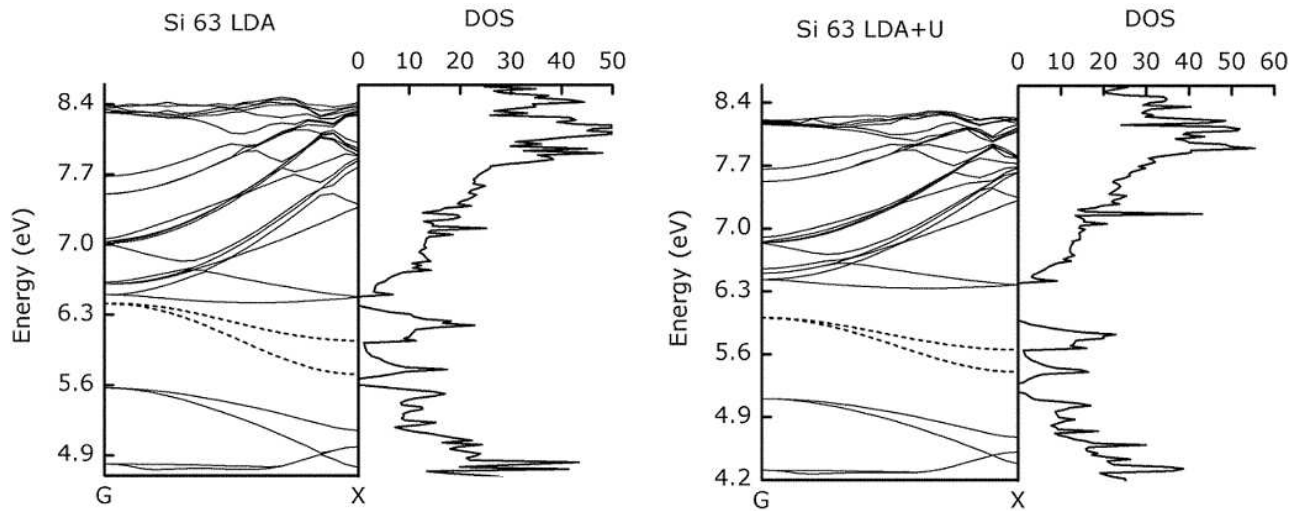}
\caption{\label{fig:si63}{\small Band structure and density of states of a
system containing a vacancy computed using the LDA (left) and the LDA+U
(right) methods.  The defect states are shown as dashed lines.  The
(indirect) band gap using the LDA (LDA+U) is 0.88 (1.21) eV and the vacancy
dispersion using the LDA (LDA+U) is 0.77 (0.71) eV. The bottom vacancy
state is doubly degenerate.}}
\end{figure*}

There is an inward relaxation of the
atoms around the vacancy site.  The symmetry of the vacancy determines the
impurity states' symmetry. The singlet s-state is located deep inside the
valence band (not shown in Fig. \ref{fig:si63}) while partly occupied triplet
p-states are located in the 
bandgap (Fig.\ref{fig:si63}). There is a splitting of p- states into twofold degenerate lower
band and a single upper band (almost unoccupied). There are a few major differences
between the LDA results and the LDA+U results. First, there is the
expected increase in 
the bandgap in the LDA+U compared to the LDA leading to the separation
between the vacancy states and valence and 
conduction bands. Second, the dispersion of vacancy p-states is
smaller in the LDA+U calculation. Thus, these states are more localized in
the LDA+U calculation. The 
dispersion cannot be removed in the supercell geometry completely, but the
LDA definitely seems to overestimate the itinerancy of electrons in the 
vacancy states. There is no noticeable Jahn-Teller distortion of the atoms
surrounding the vacancy in our 64-atom supercell geometry. This is
consistent with the observation by Zywietz {\em et. al.} \cite{zywi98a}
who suggest that a 128-atom supercell is required to observe the
distortion. The formation and the migration energy for the vacancy defect
from LDA+U calculations are 4.59 eV and 0.42 eV respectively. This is in
good agreement with experimental results \cite{ural99b,watk00a}

\subsubsection{\label{sec:hexintresult}Hexagonal interstitial}
\begin{figure*}
\includegraphics[width=17cm]{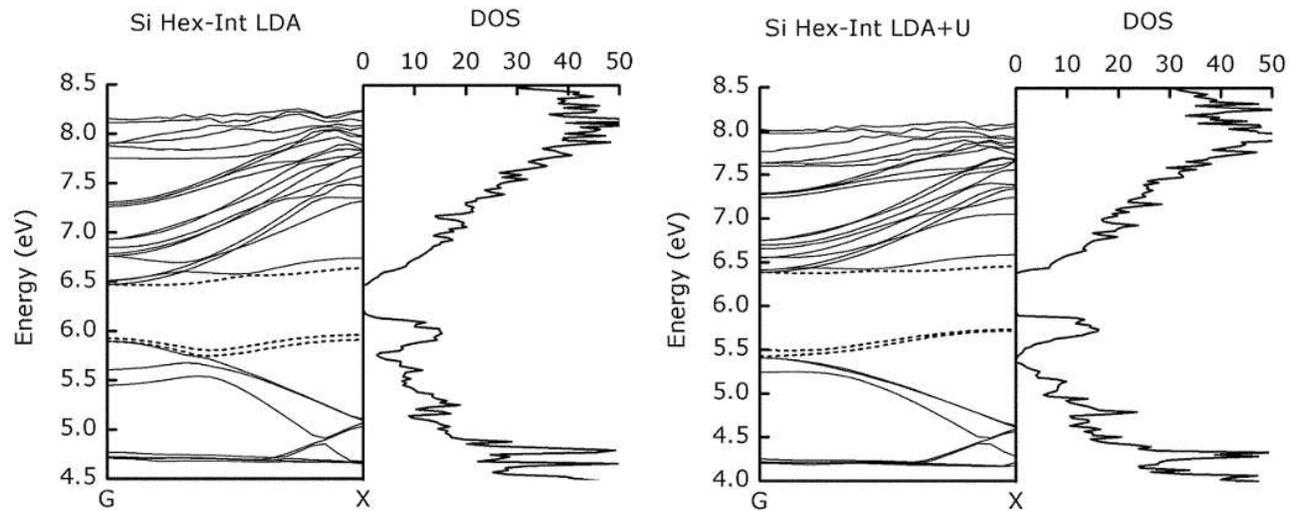}
\caption{\label{fig:si65h}{\small Band structure and density of states of a
system containing a hexagonal interstitial computed using the LDA (left)
and the LDA+U
(right) methods.  The defect states are shown as dashed lines.  The
(indirect) band gap using the LDA (LDA+U) is 0.70 (1.01) eV.
The amount of
mixing of the defect states with the valence states drops in the LDA+U
method, signifying lesser binding.}}
\end{figure*}

The self interstitial sites provide
four additional electrons to the system and 
create similar ``dangling'' bonds. Like the vacancy, they create an
almost doubly degenerate band near the bottom of the band gap and a single
band near the top of the band gap (Fig. \ref{fig:si65h}). The occupied defect states are very
close to the bottom of the band gap, in agreement with the observation made by
Needs \cite{need99a} about the defect states being quite shallow.
We find that the atoms surrounding the
interstitial move outwards in contrast to the inward movement reported by
Needs \cite{need99a}.  This is probably because, while we have optimized
the lattice constant of the system with the defect, Needs has maintained
the lattice constant at the experimental value. We suspect that Needs' approach might
cause an unphysical restriction on the relaxation. In addition to opening
the band gap, the 
introduction of the on-site repulsion quite dramatically reduces the
mixing of the 
defect states with the valence states. 
The formation energy
for the hexagonal interstitial using the LDA+U method is 4.61 eV which is
in good agreement with 
experimental \cite{ural99a} and quantum Monte Carlo \cite{leun99a} results.

\subsubsection{\label{sec:spltintresult}Split $\langle 110 \rangle$ interstitial}
\begin{figure*}
\includegraphics[width=17cm]{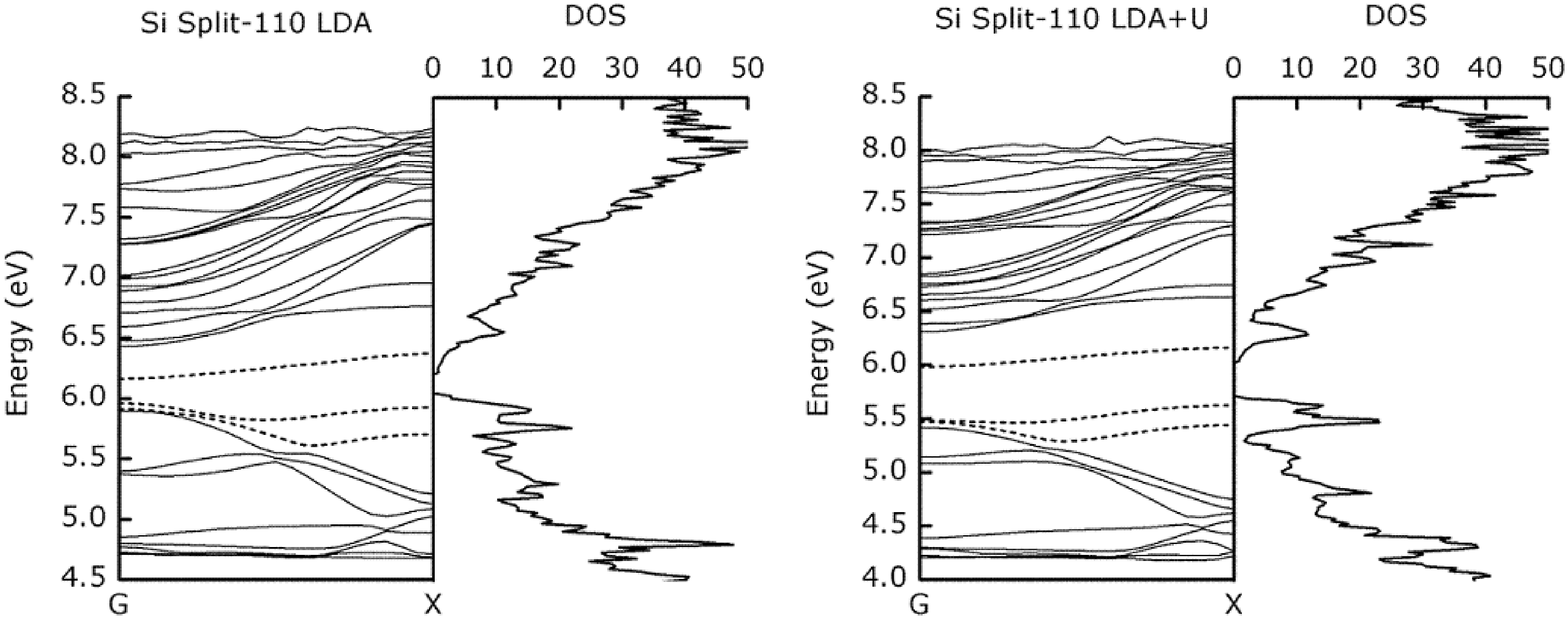}
\caption{\label{fig:si65s}{\small Band structure and density of states of a
system containing a split $\langle 110 \rangle$ interstitial computed
using the LDA (left) and the LDA+U
(right) methods.  The defect states are shown as dashed lines.  The
(indirect) band gap using the LDA (LDA+U) is 0.99 (1.12) eV.
The amount of mixing of the defect states with the valence
states drops in the LDA+U method signifying lesser binding.}}
\end{figure*}

The split
$\langle 110 \rangle$ interstitial creates defect states in the band gap
similar to the hexagonal interstitial excpet that there is a greater
splitting of the two lower bands (Fig. \ref{fig:si65s}). We attribute this to the lower symmetry
of the split $\langle 110 \rangle$ interstitial compared to the hexagonal
interstitial. Another difference is the location of the higher defect
state. It is not as close to the top as in the case of hexagonal
interstitial. We do not have a simple explanation for this. The
introduction of the on-site repulsion has a similar effect with respect to
reducing the mixing of the defect states with the bulk states. The defect
formation energy using the LDA+U method is 4.70 eV which is in good
agreement with the experimental \cite{ural99a} and quantum Monte Carlo
\cite{leun99a} results. The migration energy from the hexagonal to the
split $\langle 110 \rangle$ configurations of the interstitial was
computed using the nudged elastic band menthod \cite{NEB}. We get a value
of 0.44 eV using the LDA+U method. We are not aware of any experimental
result for this specific value.

\subsection{\label{sec:selfdiffresult}Self diffusion in Si}

The contribution of the mechanism to the diffusion process is determined
by the activation energy and the prefactor. Because all the three
mechanisms considered in this article (viz., vacancy, interstitial, 
and concerted exchange) involve similar number of atoms, one can, to a first
approximation, assume that all of them have similar entropic effects and
hence similar prefactors. Thus, based on our calculations of
the activation energies using the LDA+U method, the vacancy mechanism 
[which has an activation energy of 5.01 eV (sum of formation (4.59 eV) 
and migration (0.42 eV) energies)]
and the hexagonal to split $\langle 110 \rangle$ interstitialcy mechanism
[which has an activation energy of 5.14 eV (sum of formation (4.7 eV) and
migration (0.44 eV) energies)]
are quite close in activation energy (within the error of the method), and
should contribute equally to self-diffusion. The concerted exchange
mechanism (which has an activation energy of 5.82 eV) is less significant 
in this respect. Experimental results give a
similar indication \cite{ural99a,fahe89b}.

\section{\label{sec:summary}Summary}
The present work identified the cause for the poor description of point
defect energetics by the LDA. It corrected for the deficiencies of the LDA
by using the LDA+U method. This gave much better agreement of the
calculated activation energies with experimental observations. This method
can therefore be used for better description of diffusion in similar
semiconductor materials.

\begin{acknowledgments}
This work was funded by the United States Department of Energy, Basic
Energy Sciences Grant No. DE-FG03-99ER45788.  The computations were
performed on the NPACI IBM p690, NERSC IBM SP RS/6000 (as part CMSN),
and on the PC
clusters of the Multiscale Simulation Laboratory, Stanford: {\sc multipod}
and {\sc xepod}. Grants from NPACI and NERSC are sincerely appreciated.
\end{acknowledgments}

\bibliography{ICAM2003lanl}

\end{document}